\documentclass[12pt]{iopart}

\usepackage{iopams}  
\usepackage{graphicx}
\graphicspath{ {./images/} }
\usepackage[colorlinks]{hyperref}
\begin{document}

\title[]{Generation of a time-bin Greenberger--Horne--Zeilinger state with an optical switch}

\author{Hsin-Pin Lo*, Takuya Ikuta, Koji Azuma, Toshimori Honjo, William J. Munro, and Hiroki Takesue}

\address{NTT Basic Research Laboratories, NTT Corporation, 3-1 Morinosato Wakamiya, Atsugi, Kanagawa, 243-0198, Japan}
\ead{hsinpin.lo.cn@hco.ntt.co.jp}
\vspace{10pt}
\begin{indented}
\item[]
\end{indented}

\begin{abstract}
Multipartite entanglement is a critical resource in quantum information processing that exhibits much richer phenomenon and stronger correlations than in bipartite systems. This advantage is also reflected in its multi-user applications. Although many demonstrations have used photonic polarization qubits, polarization-mode dispersion confines the transmission of photonic polarization qubits through an optical fiber. Consequently, time-bin qubits have a particularly important role to play in quantum communication systems. Here, we generate a three-photon time-bin Greenberger--Horne--Zeilinger (GHZ) state using a 2$\times$2 optical switch as a time-dependent beam splitter to entangle time-bin Bell states from a spontaneous parametric down-conversion source and a weak coherent pulse. To characterize the three-photon time-bin GHZ state, we performed measurement estimation, showed a violation of the Mermin inequality, and used quantum state tomography to fully reconstruct a density matrix, which shows a state fidelity exceeding 70$\%$. We expect that our three-photon time-bin GHZ state can be used for long-distance multi-user quantum communication.
\end{abstract}

%

\section{Introduction}
Quantum entanglement \cite{horodecki2009quantum} is a fundamental resource in quantum information processing (QIP) and especially quantum computation \cite{nielsen2002quantum}. Such entanglement has been generated in many experiments using different physical systems including ion traps \cite{steane1996ion}, superconducting circuits \cite{steffen2006measurement}, quantum dots \cite{hanson2007spins}, and photonics \cite{kwiat1995new}. Photonic qubits based on the polarization \cite{kwiat1995new}, temporal \cite{takesue2009implementation}, and frequency modes \cite{lukens2017frequency} are suitable for information transmission over long distances. There have been many photonic QIP demonstrations using bipartite entangled states, which include quantum key distribution \cite{gisin2002quantum}, quantum communication \cite{brendel1999pulsed,gisin2007quantum}, quantum teleportation \cite{bouwmeester1997experimental}, and entanglement swapping \cite{jennewein2001experimental}. Recent loophole-free tests of Bell inequality show how different the quantum world is from the classical one \cite{hensen2015loophole,shalm2015strong,giustina2015significant}. 
\begin{figure*}[t]
\centering 
\includegraphics[width=15cm]{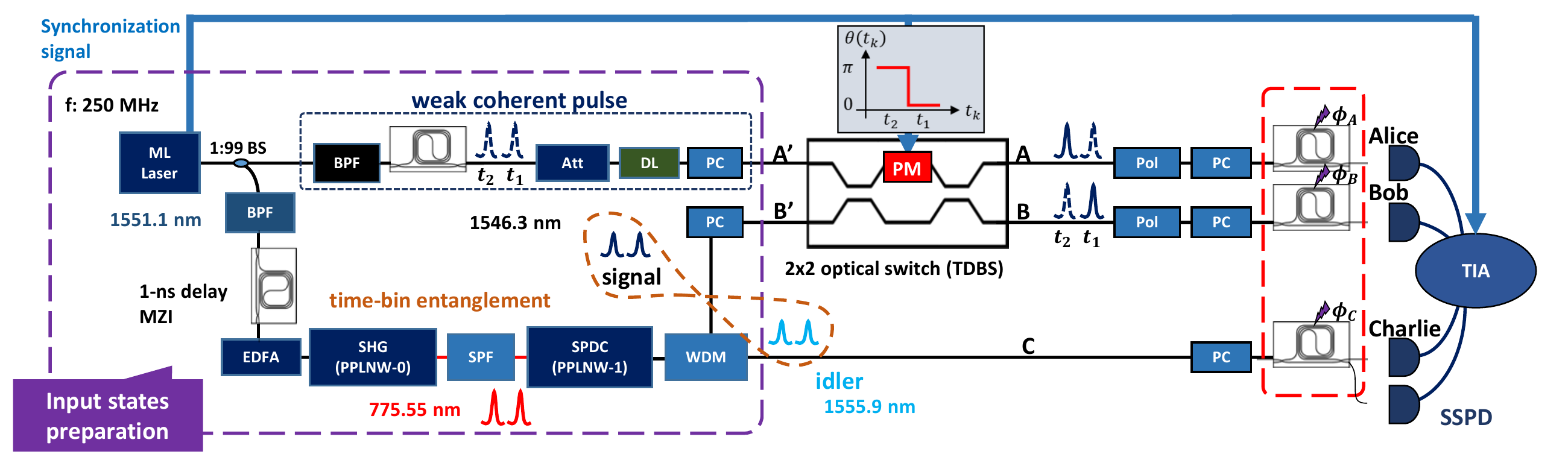} 
\caption{Experimental setup for three-photon time-bin GHZ state generation and measurement using an SPDC source and a weak coherent pulse. ML laser: mode-locked laser. f: repetition rate. BS: beam splitter. EDFA: erbium-doped fiber amplifier. PPLNW: periodically poled lithium niobate waveguide. Att: optical attenuator. DL: optical delay line. BPF: bandpass filter. SPF: short-pass filter. SHG: second harmonic generation. WDM: wavelength division multiplexing. PM: phase modulator. PC: polarization controller. Pol: polarizer. SSPD: superconducting single-photon detector. TIA: time-interval analyzer.}
\label{fig1}
\end{figure*}

Multiphoton Greenberger--Horne--Zeilinger (GHZ) states \cite{Greenberger1989going,Pan2012multiphoton} were realized more than two decades ago by using three-photon polarization qubits \cite{bouwmeester1999observation,Pan2000experimental}, and more recently they have been achieved with more photons \cite{wang2016experimental,chen2017observation,zhong201812}. These achievements were based on entangling polarization qubits with polarization beam splitters (PBSs), which is now a standard method to generate polarization GHZ states. Reported applications of multi-photon GHZ states include quantum secret sharing \cite{hillery1999quantum,tittel2001experimental,xiao2004efficient}, quantum error-correction \cite{bell2014experimental}, clock synchronization \cite{komar2014quantum}, boson sampling \cite{zhong201812}, and all-photonic quantum repeaters \cite{azuma2015all,hasegawa2019experimental,li2019experimental,zhang2022loss}. Multi-user quantum conference key agreement has also been implemented \cite{grasselli2018finite,proietti2021experimental,thalacker2021anonymous}, which also showed advantages over multi-user conference key distribution with bipartite entanglement \cite{epping2017multi}. The above investigations are all important for quantum networks \cite{kimble2008quantum} using a multiphoton entanglement.

For a large-scale quantum network, we need robust photonic qubits that can be transmitted and processed over long distances for multiple users. However, polarization states of photons are generally perturbed by polarization mode dispersion (PMD) in optical fibers, implying that a photonic polarization qubit is unsuitable for fiber-based quantum communication. A time-bin photonic qubit can carry quantum information through an optical fiber without the effect of PMD, which is an advantage over polarization qubits \cite{Honjo2008long,takesue2015quantum}. Recently, we have implemented time-bin qubit quantum logic gates \cite{takesue2014entangling,lo2018entanglement,lo2020quantum} using a high-speed two-input two-output (2$\times$2) optical switch as a time-dependent beam splitter (TDBS) \cite{takesue2014entangling}. These gates can be used for essential gate operations on time-bin qubits required for quantum computing and communication related tasks. 

Our next challenge is to generate multiphoton time-bin entangled states. Our particular interest is in generating time-bin photonic GHZ states, which have versatile quantum computation and communication applications \cite{hillery1999quantum,epping2017multi} and are exceptionally useful building blocks for realizing more complex quantum states \cite{briegel2001persistent,browne2005resource}. Here, as a device analogous to a PBS for generating a polarization GHZ state, we used the TDBS to entangle time-bin entangled photon pairs prepared by a spontaneous parametric down-conversion (SPDC) process and a weak coherent pulse (WCP) \cite{rarity1999three}. We then performed measurement estimations of four joint measurements to confirm our state violated the Mermin inequality \cite{mermin1990extreme}. Then, we used quantum state tomography (QST) \cite{james2001measurement} to reconstruct the density matrix of the tripartite state and evaluated various important physical properties.

\section{Three-photon time-bin GHZ state}
SPDC photon pairs are the main workhorse for generating multiphoton polarization-entangled states using PBSs  \cite{bouwmeester1999observation, Pan2000experimental,wang2016experimental,chen2017observation,zhong201812,zhao2003experimental,zhao2004experimental,lu2008experimental}. Further 
a WCP can effectively improve the generation efficiency of multiphoton entanglement \cite{zhao2004experimental,lu2008experimental,mikami2005new}. However, these techniques cannot be directly applied to generate multiphoton states based on time-bin qubits. As the first experimental demonstration of a multiphoton time-bin GHZ state, we prepared a 
time-bin state $\left|+ \right\rangle_{A'}=1/\sqrt{2}(\left|t_1 \right\rangle_{A'} + \left|t_2 \right\rangle_{A'})$ from a WCP and an entangled photon pair $\left|\Phi^{+} \right\rangle_{B'C}=1/\sqrt{2}(\left|t_1t_1 \right\rangle_{B'C} + \left|t_2t_2 \right\rangle_{B'C})$ from the SPDC process. Here, time-bin states ($\left|t_1\right\rangle, \left|t_2 \right\rangle$) represent a photon in two different temporal modes used to encode the computational-basis states $\left|0 \right\rangle$ and $\left|1 \right\rangle$ \cite{brendel1999pulsed}, while the subscript labels represent spatial modes. Our initial state can then be described by $\left|\Psi \right\rangle_{A'B'C}=\left|+ \right\rangle_{A'} \otimes \left|\Phi^{+} \right\rangle_{B'C}$. We employed the TDBS \cite{takesue2014entangling} to entangle two time-bin states from independent light sources (one from the WCP and one from the SPDC), which have directly passed $\left|t_{1}\right\rangle$ (port $A'$ to $A$ and $B'$ to $B$) and reflected $\left|t_{2}\right\rangle$ (port $A'$ to $B$ and $B'$ to $A$) as shown on the right in Fig.~\ref{fig1}. This action of the TDBS for time-bin qubits corresponds to that of a PBS for polarization qubits. When photons in $A'$ and $B'$ spatial modes arrive at the TDBS at the same time
, our post-selective three-photon time-bin GHZ state is generated with the form $\left|\mbox{GHZ} \right\rangle_{ABC}=1/\sqrt{2}(\left|t_1t_1t_1 \right\rangle_{ABC} - \left|t_2t_2t_2 \right\rangle_{ABC})$ \cite{zhao2004experimental,lu2008experimental}. The minus sign between the two terms in the $\left|\mbox{GHZ} \right\rangle_{ABC}$ state is due to $\left|t_2 \right\rangle$ states passing through TDBS. 

\section{Experimental setup}
Our experiment setup is depicted in Fig.~\ref{fig1}. We prepared a 1551.1-nm pulse train by launching mode-locked (ML) laser pulses with a repetition rate of 250 MHz into a 0.2-nm optical bandpass filter (BPF). The single-pulse train is then launched into a 1-bit delay unbalanced Mach-Zehnder interferometer (UMZI) to prepare double pulses with a 1-ns time interval. The double pulses are then amplified by an erbium-doped fiber amplifier (EDFA) and directed into a periodically poled lithium niobate waveguide (PPLNW-0) to prepare 775.55-nm double pulses using second harmonic generation (SHG). A short-pass filter (SPF) is then used to remove the input pump light. The SHG light is then input into another PPLN waveguide (PPLNW-1), where the 1.5-$\mu$m time-bin entangled photon pairs $\left|\Phi^{+} \right\rangle_{B'C}$ are generated via spontaneous parametric down-conversion. The signal (1546.3 nm) and idler (1555.9 nm) photons are separated by a wavelength division multiplexing (WDM) filter with a 0.2-nm bandwidth for both the signal and idler channels, whose average photon number per pulse is $\sim$0.0082. Meanwhile, we use a small fraction of ML lasers sent to another BPF as a WCP source with the same central wavelength and bandwidth as the signal photons of the SPDC photon pair. The pulses are then directed into another UMZI to generate the double pulses of WCP as a single time-bin state. At the same time, we set the relative phase of UMZI to prepare the $\left|+\right\rangle_{A'}$ state.

\begin{figure}[t]
\centering 
\includegraphics[width=15cm]{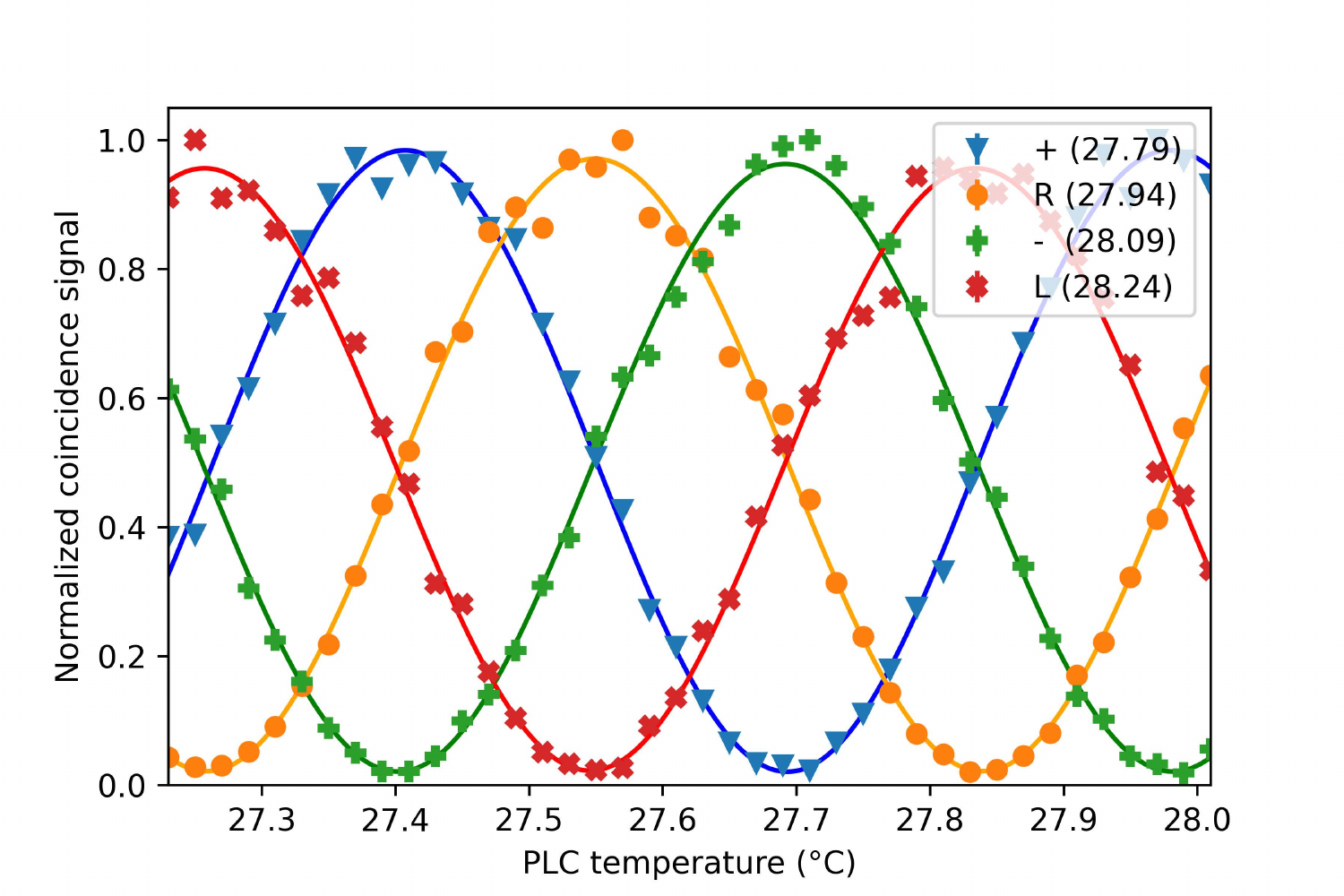}
\caption{Two-photon interference measurements between Bob and Charlie. Normalized coincidence signal observed as a function of Charlie’s relative phase ($\phi_C$) while fixing Bob's relative phase ($\phi_B$) as $\{\left|+ \right\rangle$, $\left|R \right\rangle$, $\left|- \right\rangle$, and $\left|L \right\rangle \}$. The relative phase is set by changing base plate temperature of their PLCs, respectively. The solid lines are the fitting results.
}
\label{fig2-1}
\end{figure}

The prepared double pulse of WCP and the time-bin entangled state are directed into input ports $A'$ and $B'$ of the 2$\times$2 optical switch, respectively. The 2$\times$2 optical switch can operate as a TDBS by adjusting the DC bias and RF modulation signal with an electric pulse of 4-ns width applied to the phase modulator (PM) in one of its optical paths \cite{takesue2014entangling}. As a result, we can form GHZ-type entanglement among a photon in the WCP and the time-bin entangled photon pair. Photons output from ports $A$ and $B$ of the switch as well as the idler photon are directed into respective UMZIs followed by a superconducting single-photon detector (SSPD) for projective measurements on time-bin qubits. The detection signals from the SSPD are input into a time-interval analyzer (TIA) for three-fold coincidence measurements. We use polarization controllers (PC) and polarizers (Pol) to eliminate polarization distinguishability of photon pairs. The insertion losses of the UMZI and the optical switch are approximately 2.0 and 3.5 dB, respectively. The UMZIs are fabricated using planar light-wave circuit (PLC) technologies \cite{takesue2005generation}. We can precisely control the relative phase of UMZI by adjusting the base plate temperature of the PLC. To reduce the number of projective measurements, we use both output ports of Charlie's UMZI, where the phase difference between his two outputs is $\pi$. The detection efficiencies of SSPDs for Alice and Bob are 57$\%$ and 52$\%$, respectively, and those of two SSPDs for Charlie are 62$\%$ and 46$\%$. The dark count rates of SSPD are all lower than 40 cps. The three-fold coincidence count rate is approximately 0.8 count/min. We did not subtract accidental coincidences in any of the experiments reported here.

\begin{figure}[t]
\centering 
\includegraphics[width=15cm]{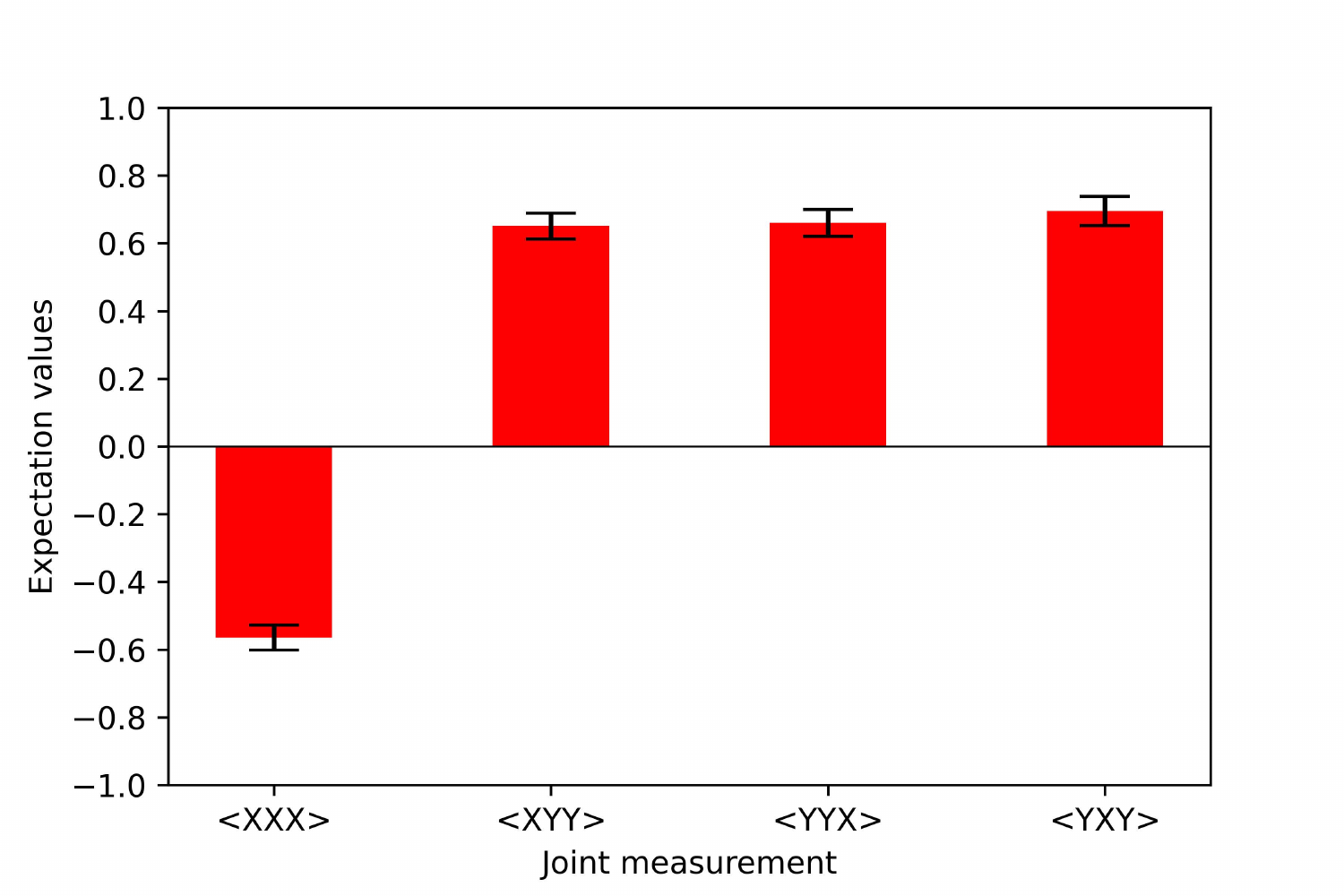}
\caption{Expectation values and error estimates for the four joint measurements $\langle XXX \rangle$,  $\langle XYY \rangle$, $\langle YYX \rangle$, and $\langle YXY \rangle$ required to test the Mermin inequality. 
}
\label{fig3}
\end{figure}

\section{Experimental results}
\subsection{Time-bin entangled states measurement}
It is important to first confirm that Bob and Charlie have generated two-qubit entangled states $\left|\Phi^{+} \right\rangle_{B'C}$. We turn off the RF signal to the PM, so that the 2$\times$2 optical switch works like a tunable beam splitter, where the splitting ratio can be adjusted using DC bias. Here, we adjust the DC bias so that all photons enter Bob's port $B$. We set the temperature of their UMZIs so that each of their single-photon detection projects the received time-bin qubit into state $(|t_1 \rangle+e^{i \phi_j} | t_2 \rangle )/\sqrt{2}$ with $j \in \{A,B,C\}$, as shown in Fig.~\ref{fig1}. Then, to confirm the entanglement, Bob fixed his measurement setup so as to project his received qubit into states $\{|+\rangle, |R\rangle, |-\rangle, |L \rangle \}$ for each of which Charlie performs measurement with changing the temperature of his PLC, where $\left|\pm \right\rangle=(\left|t_1 \right\rangle \pm \left|t_2 \right\rangle)/\sqrt{2}$, $\left|R \right\rangle=(\left|t_1 \right\rangle +i \left|t_2 \right\rangle)/\sqrt{2}$ and $\left|L \right\rangle=(\left|t_1 \right\rangle -i \left|t_2 \right\rangle)/\sqrt{2}$. The relative phase $\phi_j$ changes proportionally with the PLC's base plate temperature, and it shifts $\pi$ radians for the temperature difference of 0.3$^\circ$C. The normalized coincidence signals are shown in Fig.~\ref{fig2-1}. The average visibility was 94.7$\pm$0.94$\%$ \cite{sdpcvisibility}. Our state can also violate the CHSH-Bell inequality with the expectation value $S$=2.65$\pm$0.019 of the Bell operator by more than 34 standard deviations \cite{hong1987measurement,lo2011beamlike}. Thus, we confirmed that the prepared state is close to a pure time-bin maximally entangled state. It is now time to turn our attention to the generation and characterization of the GHZ state.

\subsection{Joint measurements and Mermin inequality}
With the generation of our GHZ state, it is critical to confirm that we have created a tripartite entangled state. The simplest way to achieve this with the fewest measurements is to use the Mermin operator \cite{mermin1990extreme} defined by 
\begin{eqnarray}
M = X_A Y_B Y_C+ Y_A X_B Y_C + Y_A Y_B X_C - X_A X_B X_C
\end{eqnarray}
with $X_i$ and $Y_i$ being the usual Pauli operators for the $i{\mbox{th}}$ photon, which can be measured in the $+/-$ and $R/L$ basis in our setting, respectively. It is well known that local realism follows $| \langle M\rangle| \leq 2$ \cite{seevinck2001sufficient,toth2005addendum}. Quantum mechanically $\vert \langle M \rangle \vert \leq 4$, meaning there is a region of $ 2<\vert \langle M \rangle \vert \leq 4$ where the quantum and classical predictions differ (this is similar to the case of $S \in (2, 2 \sqrt{2}]$ in the bipartite scenario with a CHSH-Bell inequality). A measurement of $\vert \langle M\rangle \vert >2$ clearly indicates the existence of multipartite GHZ entanglement. To achieve this, we need to measure the four expectation values $\langle X_A Y_B Y_C \rangle$, $\langle Y_A X_B Y_C \rangle$, $\langle Y_A  Y_B X_C \rangle$, and $\langle X_A X_B X_C \rangle$, which are shown in Fig.~\ref{fig3}. For example, we can determine the $\langle Y_A  Y_B X_C \rangle$ expectation value when Alice and Bob measure their photons in the $R/L$ basis, while Charlie measures in the $+/-$ basis by setting the relative phase $\phi_{j}$ of their PLCs \cite{sdpcvisibility}, respectively. Those values allow us to estimate $\vert \langle M \rangle \vert=2.57\pm0.04$ \cite{jointmeasurementtimebin}, which indicates that our three-photon time-bin state violates local realism by more than 14 standard deviations. This confirms the presence of entanglement among the three photons \cite{mermin1990extreme,toth2005addendum}. 

\begin{figure*}[t] 
\centering 
\includegraphics[width=15cm]{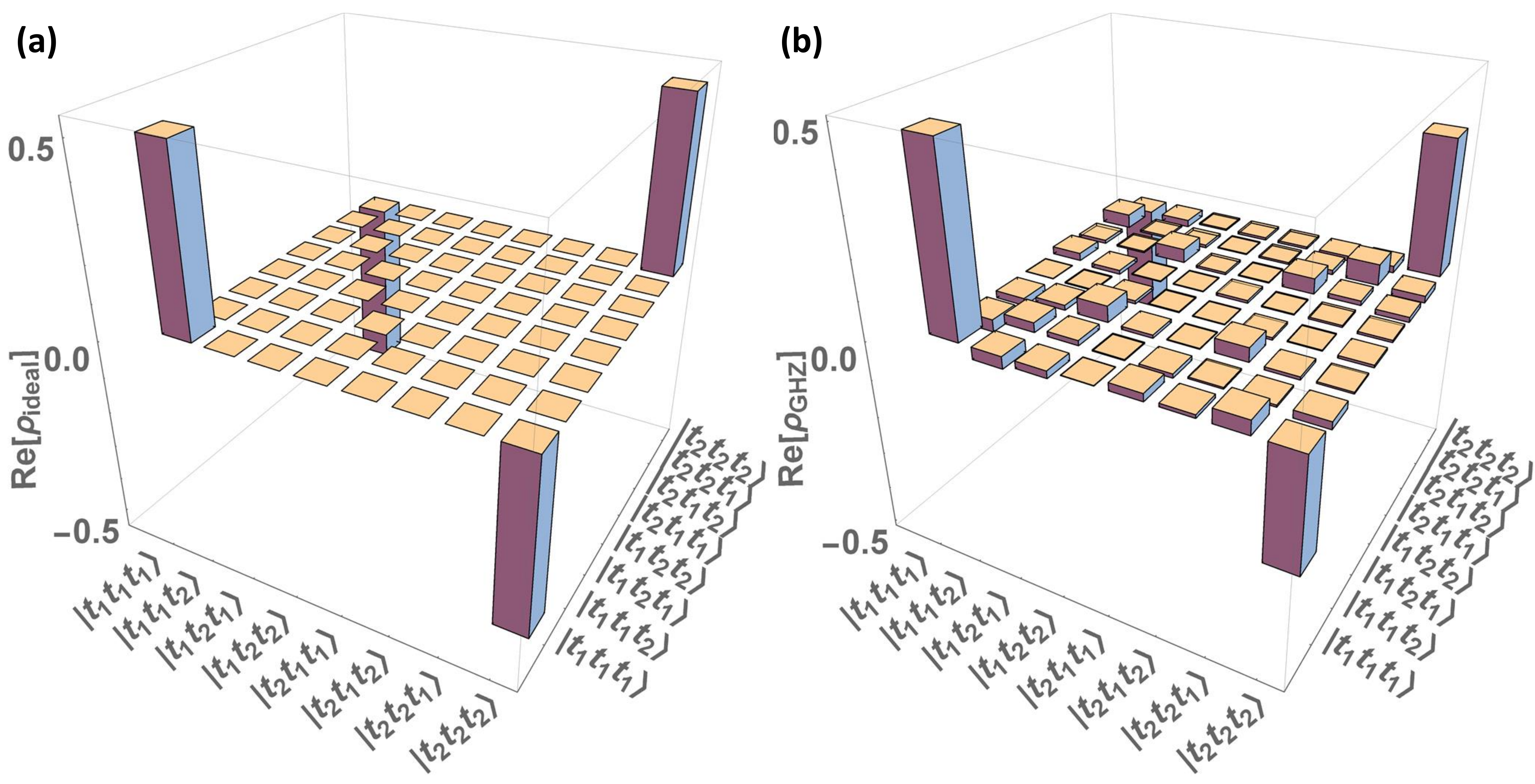}   
\caption{Reconstruction of the density matrix of three-photon time-bin GHZ states using quantum state tomography. (a) The ideal density matrix ($\rho_{\mbox{ideal}}$) of $\left|\mbox{GHZ} \right\rangle$. (b) The reconstructed density matrix ($\rho_{\mbox{GHZ}}$), which includes maximum-likelihood estimation. The imaginary part of the density matrix is small (approximately zero), so it is not shown.} 
\label{fig2}
\end{figure*}

\subsection{Quantum state tomography of a three-photon time-bin GHZ state}
To fully characterize our three-photon time-bin states, we performed quantum state tomography measurement on them \cite{takesue2009implementation,james2001measurement,resch2005full}. By setting the relative phase $\phi_j$ of the UMZIs in front of the SSPDs, we can implement all four projective measurements onto states $\{\left|t_1 \right\rangle$, $\left|t_2 \right\rangle$, $\left|+ \right\rangle,$ and $\left|R \right\rangle \}$. By implementing a total of 64 projective measurements, we were able to reconstruct the density matrix of those three-photon states. To obtain the physical density matrix, we performed maximum-likelihood estimation \cite{james2001measurement}, which allows us to reconstruct the physical density matrix ($\rho_{\mbox{GHZ}}$) of our three-photon time-bin GHZ state as shown in Fig.~\ref{fig2}. By comparing the measured state to the ideal GHZ state, we observed a state fidelity of $F(\rho_{\mbox{GHZ}})=\left\langle \mbox{GHZ}\left|\rho_{\mbox{GHZ}} \right|\mbox{GHZ} \right\rangle_{ABC}=0.713\pm0.068$ was observed. 

\begin{table}[!ht]
\centering
\begin{tabular}{c| c} 
 \hline
 State fidelity & 0.713 $\pm$ 0.068 \\ 
 \hline
 Linear entropy & 0.49 $\pm 0.016$ \\ 
 \hline
 Von Neumann entropy & 1.198 $\pm$ 0.148 \\
 \hline
 Purity & 0.571 $\pm$ 0.068 \\
 \hline
  Tripartite negativity 
   & 0.59 $\pm$ 0.065 \\
 \hline
 Entanglement witness & -0.213 $\pm$ 0.068 \\
 \hline
\end{tabular}
\caption{Experimentally relevant quantities derived from our experimental density matrix.}
\label{table}
\end{table}
We can use this reconstructed density matrix to determine experimentally relevant physical quantities, including the degree of entanglement, linear entropy, von Neumann entropy, and purity, as shown in Table \ref{table}. To obtain a measure of tripartite entanglement valid for non-pure states, we estimate the tripartite negativity $N_{ABC}(\rho)$ of the three-qubit state $\rho$. The tripartite negativity \cite{sabin2008classification} is defined as $N_{ABC}(\rho)=(N_{A-BC}N_{B-AC}N_{C-AB})^{1/3}$, where the bipartite negativities are defined as $N_{P-JK} =-2\sum_{i} \sigma_{i}(\rho^{TI})$ with $\sigma_{i}(\rho^{TI})$ being the negative eigenvalues of the partial transpose of $\rho$ with respect to the subsystem, where $P=A,B,C$ and $JK=BC, AC, AB$, respectively. If any of the bipartite negativities are zero, then the tripartite negativity is also zero. Our state has a non-zero tripartite negativity of 0.59 $\pm$ 0.065 \cite{sabin2008classification,buscemi2011measure,3qubitsNegativityntabc}, clearly indicating tripartite entanglement. $N_{ABC}(\rho)>0$ is a sufficient condition for distillability to a GHZ state (GHZ-distillability) \cite{sabin2008classification,dur1999separability}. 

We also characterized our state using an entanglement witness operator ($\mathcal{W}$) to detect the GHZ entanglement \cite{acin2001classification}, $\mathcal{W}=I/2-\left|\mbox{GHZ} \right\rangle \left\langle \mbox{GHZ}\right|$, where $I$ is the identity matrix. When the expectation value of this witness operator is negative, the state implies the existence of genuine GHZ entanglement. The expectation value is $\langle\mathcal{W}\rangle=-0.213\pm0.068$, which is negative by over three standard deviations. This clearly confirms that we successfully generated a three-photon time-bin GHZ state. 



\section{Discussion and conclusion}

The current three-fold coincidence rate is still low and the fidelity is limited to 71$\%$. Such limited performance in terms of the generation rate and the fidelity is mainly attributed to the purity-heralded efficiency trade-off observed in a spectrally filtered single photons from frequency-correlated photon pair sources \cite{meyer2017limits,blay2017effects}. 
Significant improvements in generation rate and fidelity are expected by using a source of photon pairs that generate separable states \cite{grice2001eliminating,kaneda2016heralded,graffitti2018design} and reducing losses in optical components.  

In conclusion, we have experimentally generated a three-photon time-bin GHZ state using time-bin Bell states from SPDC photon pairs and 
a weak coherent pulse interacting via a 2$\times$2 optical switch. We performed an estimation of measurements and showed that the resulting entangled states directly violate the Mermin inequality. Furthermore, we performed quantum state tomography to reconstruct the density matrix of our generated state, from which we could determine a state fidelity greater than 70$\%$ as well as a tripartite entanglement negativity of $\sim$0.6. From these experimental results, it is clear that our state possesses tripartite entanglement of the GHZ form. We expect this technology to be used for multipartite quantum key distribution \cite{proietti2021experimental,thalacker2021anonymous,chen2004multi}, quantum secret sharing \cite{hillery1999quantum,tittel2001experimental,xiao2004efficient}, quantum repeaters \cite{azuma2015all}, and distributed quantum computation through optical fiber for a large-scale quantum network \cite{kimble2008quantum}.


\section*{References}

\bibliographystyle{iopart-num}
\bibliography{main}

\begin{thebibliography}{10}
\bibitem{sdpcvisibility}
For the two-photon interference measurements, we set $\phi_{B}$ at 0
  (27.79$^\circ$C, $\left|+ \right\rangle=(\left|t_1 \right\rangle + \left|t_2
  \right\rangle)/\sqrt{2}$, $V_{\left|+ \right\rangle}=96.36\pm0.89\%$),
  $\pi/2$ (27.94$^\circ$C, $\left|R \right\rangle=(\left|t_1 \right\rangle +i
  \left|t_2 \right\rangle)/\sqrt{2}$, $V_{\left|R
  \right\rangle}=93.41\pm1.05\%$), $\pi$ (28.09$^\circ$C, $\left|-
  \right\rangle=(\left|t_1 \right\rangle - \left|t_2 \right\rangle)/\sqrt{2}$,
  $V_{\left|- \right\rangle}=94.24\pm0.94\%$), and $3\pi/2$ (28.24$^\circ$C,
  $\left|L \right\rangle=(\left|t_1 \right\rangle -i \left|t_2
  \right\rangle)/\sqrt{2}$, $V_{\left|L \right\rangle}=94.99\pm0.89\%$), where
  $V$ is the visibility of each measurements.
\bibitem{jointmeasurementtimebin}
Expectation values for the four joint measurements $\langle XXX \rangle$,
  $\langle XYY \rangle$, $\langle YYX \rangle$, and $\langle YXY \rangle$ are
  -0.565$\pm$0.038, 0.651$\pm$0.038, 0.660$\pm$0.041, and 0.695$\pm$0.043,
  respectively.
\end{thebibliography}

\end{document}